# SOCIAL ASPECTS OF VIRTUAL TEAMS

Daphna Shwarts-Asher
The College for academic Studies, Hayotzrim 2, Or Yehuda, Israel
e-mail: daphna@mla.ac.il

## ABSTRACT

There has been a transformation from individual work to team work in the last few decades (Ilgen, 1999), and many organizations use teams for many activities done by individuals in the past (Boyett & Conn, 1992 ; Katzenbach & Smith, 1993). In recent years, there has been a renewed interest in computer-mediated groups because of the increases in globalization of business operations leading to geographically dispersed executives and decision makers. However, what seems to be lacking is some focus in terms of problem settings and corresponding tools to support collaborative decision making. The research question of this study deals with the dynamics of virtual teams' members. A model, suggesting that team dynamics can increase the teams' output, is presented, and a methodology to examine the model is illustrated. An experiment was performed, in which subjects, who were grouped into teams, had to share information in order to complete a task. The findings indicate that the social aspect of the virtual team's discussion is negative than the social aspect of the face-to-face team's discussion, and that the virtual team's output is inferior to the face-to-face team's output. The virtual team is a common way of working nowadays, and with the growing use of Internet applications and firms' globalization it will expand in the future. Thus, the importance of the theoretical and practical implementation of the research will be discussed.

## INTRODUCTION

A growing number of organizations are adopting virtual team systems to deal with a rapidly changing environment. The field of team dynamics refers to the work of traditional face-to-face teams and needs to be adjusted to the 21$^{st}$ century and its types of communication media, specifically to the virtual environment. This study attempts to provide better practices in an e-work environment and propose a theoretical model that illustrates monitoring and assessment of activity in computer-supported collaborative projects. The research question asks whether the output of virtual teams (namely teams whose members do not meet face-to-face) such as efficiency, effectiveness and satisfaction, is affected by the team members interaction. This question deals with the dynamics within the team: Is it social oriented? Or maybe task oriented? Is the social aspect of team work characterized by positive or negative phrases? Is the task aspect of team work characterized by question or answer phrases? The aim is to trace the mechanism in which output is achieved in virtual settings, and the ways that these processes differ from face-to-face settings. This research contributes to a better understanding of virtual teams in hope of improving the teams work in the virtual world.

## BALES' IPA MODEL

Hackman (1987) suggested that team design influences the team's output through its impact on processes. Chidambaram & Tung (2005) claimed that performance is driven by social demands and task demands. Bales (1950) developed a method to measure the team dynamics, called Interaction Process Analysis (IPA). Bales' IPA is a method for analyzing the "systems of human interaction" in, originally, small face-to-face groups. The heart of the method is a technique of classifying behaviors act by act, as it occurs, and a series of ways to analyze the data and obtain descriptive indices of group process. The IPA consists of 12 complementary-paired group processes; these are further subdivided into four major functions, describing communications issues or problems. Fahy (2006) found the IPA itself to be useful for describing interaction processes in online groups.

## THE RESEARCH MODEL & HYPOTHESES

The research model was developed as depicted in Figure 1. According to the model, the virtuallity level is an affecting variable and it is referred to as the input of the team. The dependent variables are the outputs of the team work: efficiency, effectiveness and satisfaction. There are three mediators, which can be described as social versus and task characteristics of the discussion: social versus task ratio, positive versus negative ratio and questions versus answers ratio.

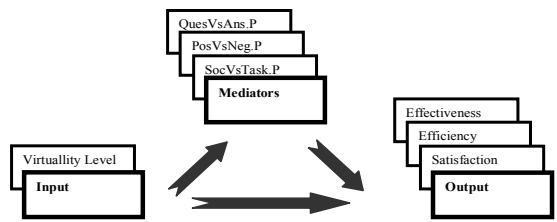

*Figure 1: The research model*

Cramton & Webber (2005) referred to the limitation of information evolves from using different types of technologies and to the increase in hostility and aggression of the communicators toward each other. Denton (2006) and Potter & Balthazard (2002) claimed that virtual teams are less successful than face-to-face teams on most outcome measures. I argue that distant communication is a potential source of complications due to a lack of tangible information. Therefore, I assume a negative correlation between virtuallity and both the mediators and the output variables.

Montoya-Weiss, Massey & Song (2001) claimed that managing virtual teams' conflict is the key to their success; Guerra et al. (2005) showed that the social conflict decreases satisfaction and well-being; Diefendorff & Gosserand (2003) stated that emotional processes at work can lead to positive or negative team output. My prediction is that the discussion orientation will positively impact team output. Using Bales (1950) Interaction Process Analysis (IPA) three mediators were defined, which indicate the discussion orientation.

Organizing the team's work should enhance its outputs. Adopting different types of technologies is one way of teams to achieve this target, yet the virtual working environment complicates the team work. Denton (2006) and Potter & Balthazard (2002) showed that virtual teams are less successful than face-to-face teams on most outcome measures. Therefore:

*Hypothesis no. 1 – for an intellective task (a task that has a correct answer), a virtual team's output is inferior to a face-to-face team's output (for the same task).*

Clausen & Sung (2008) claimed that computer-mediated (virtual) teams' social processes are different than face-to-face teams. The group of hypothesis no. 2 concerns with the impact of virtuallity on the content of the team discussion.

Saunders & Ahuja (2006) stated that in temporary teams, members are more focused on task-related outcomes. Therefore:

*Hypothesis no. 2a – for an intellective task, virtual team's discussion is task oriented, while face-to-face team's discussion is social oriented.*

Cramton & Webber (2005) argued that virtual environment increases hostility and aggression of the communicators toward each other. Thus:

*Hypothesis no. 2b – for an intellective task, the social part of the virtual team's discussion is negative oriented, while the social part of the face-to-face team's discussion is positive oriented.*

In Bales (1950) model, giving and requesting information are activities related to the group's task. The observed proportions reported by Bales, for traditional teams, were 56.4% task – giving transcript and 9.7% task – asking transcript. The observed proportions reported by Fahy (2006), for online communicators, were 70% task – giving transcript and 14% task – asking transcript. We will also assume that:

*Hypothesis no. 2c – for an intellective task, the task part of the virtual and face-to-face team's discussion is answer oriented, rather than question oriented.*

Diefendorff & Gosserand (2003) stated that emotion processes can lead to positive or negative team output. The group of hypothesis no. 3 concerns with the impact of the team discussion content, namely the social versus and task characteristics of the discussion, on the teams output.

The first hypothesis assumes the social context (versus task one) will satisfy the team members while hurting their success in achieving the task:

*Hypothesis no. 3a – for an intellective task, the Social versus Task orientation of the discussion influences the team's output: a positive influence on Satisfaction, a negative influence on Effectiveness and Efficiency.*

Ashkanasy, Hartel & Daus (2002) claimed that positive emotions lead to positive outputs, whereas negative emotions lead to negative one. I also predict that:

*Hypothesis no. 3b – for an intellective task, the Positive versus Negative orientation of the social part of the discussion influences the team's output: a positive influence on Satisfaction, Effectiveness and Efficiency.*

The last hypothesis is based on the approach that giving information and supplying answers are activities that promote the teams' task, rather than asking questions:

*Hypothesis no. 3c – for an intellective task, the Questions versus Answers orientation of the task part of the discussion influences the team's output: a negative influence on Satisfaction, Effectiveness and Efficiency.*

The next section describes the methodology used in order to test the hypotheses. The results will follow.

## METHODOLGY

An experiment was designed, in which a team task was delivered to 150 undergraduate students in an academic college. The subjects, who were grouped into teams of three members, had to share information in order to complete the task. Each team was given a task that takes approximately 30 minutes to complete. The research design includes the type of communication (virtual vs. face-to-face) using a total of two experimental conditions. The virtual condition was implemented on 26 teams, while the Face-to-face condition was implemented on 24 teams, as described in Table 1. Thus, the experiment included 150 subjects (2 conditions * 24-26 teams * 3 subjects).

|   | Virtuallity Level | N | Remarks |
|---|---|---|---|
| 1 | 0 | 24 | Face-to-face team |
| 2 | 1 | 26 | virtual team |

*Table 1: Experimental Conditions*

**Procedure**

Subjects were invited in groups of three to meetings that were conducted using MSN-Messenger (virtual) or face-to-face (non virtual) communication. The process of the experiment includes an intellective task. Each team member received a discrete and different piece of information, and only the aggregation of all the information revealed the whole "picture" and led to the correct solution.

**Operationalization of the Dependent Variables and mediators**

<u>Efficiency</u>- the time required to complete the task.
<u>Effectiveness</u>- the team's solution compared to the correct solution.
<u>Satisfaction</u>- team members' reaction to the task will be measured by their understanding of communication, and satisfaction of medium, results and process.
<u>Task and Social Processes</u>
A textual (or audio) recording was recorded for each virtual (or face-to-face) meeting. Task and social processes were measured by content analysis: The analysis, from each meeting, included the number of social positive phrases compared to the number of social negative phrases; and the numbers of task question phrases compared to the number of task answer phrases, accordingly to Bales (1950) model. In order to use reliable measurements, the phrase counting was done separately by two independent judges. The two judge analyses were compared, and in a case of different decision (concerning the phrase category), a new agreed decision was made. Three measures were calculated out of the above phrases counting:

Social Vs Task Level – social phrases percentage related to task phrases percentage, among total phrases.
Positive Vs Negative Social Level – positive phrases percentage related to negative phrases percentage, among social phrases.
Questions Vs Answers Task Level – question phrases percentage related to answer phrases percentage, among task phrases.

Table 2 presents a summery of the Means and SD's of all the mediating variables; above the experiment condition (Means and SD's by the independent variable are described in the following section).

|   | M | SD |
|---|---|---|
| SocVsTask.P | 24.50 | 9.08 |
| PosVsNeg.P | 71.71 | 15.74 |
| QuesVsAns.P | 28.64 | 5.95 |

*Table 2: Means and SD's of the Mediator Variables (Overall N=50)*

**PRELIMINARY FINDINGS**

Fifty experiments where performed among undergraduate students in an academic college. Each one of the fifty teams included three participants, a total of 150 participants. The mean age of the participants was 25.79 years old (SD = 3.94). Among the participants, there were 89 males and 61 females. No significant differences was found among the teams in mean age ($F_3 = 0.676$, $p = .568$). Breslow-Day test showed that there is no gender ratio differences among the experimental cells ($\chi^2_1 = 0.766$, $p = .381$).

An analysis of variance (ANOVA) was performed for assessing the effects of the independent variable (Virtuallity) on the processes and outputs variables. Table 3 presents a summary of the Means and SDs of all the output and mediating variables by the independent variables.

|   | Virtual team | | | Face-to-face team | | | Total | | |
|---|---|---|---|---|---|---|---|---|---|
|   | M | SD | N | M | SD | N | M | SD | N |
| Effectiveness | 77.88 | 28.37 |   | 81.23 | 26.72 |   | 79.49 | 27.36 |   |
| Efficiency | 40.23 | 11.93 |   | 20.83 | 7.98 |   | 30.92 | 14.08 |   |
| Satisfaction | 3.66 | .52 | 26 | 4.28 | .35 | 24 | 3.96 | .54 | 50 |
| SocVsTask.P | 22.90 | 10.57 |   | 26.23 | 6.94 |   | 24.50 | 9.08 |   |
| PosVsNeg.P | 66.30 | 17.55 |   | 77.58 | 11.14 |   | 71.71 | 15.74 |   |
| QuesVsAns.P | 28.80 | 6.73 |   | 28.48 | 5.11 |   | 28.64 | 5.95 |   |

*Table 3: Means and SD's of All the Output and Mediator Variables by the Independent Variable*

The ANOVA indicates, for each hypothesis respectively, that:

H1 – Virtual teams are not less successful than face-to-face teams in completing the task; yet, virtual teams are less satisfied ($F_{1,46} = 25.119$, $p < .001$, $\eta^2 = 0.353$) and take longer time in carrying out the task ($F_{1,46} = 45.933$, $p < .001$, $\eta^2 = 0.500$).

H2a – Virtual team's social versus task discussion orientation is not different than face-to-face team's social versus task discussion orientation.

H2b – The social part of the virtual team's discussion is different than the social part of the face-to-face team's discussion: The positive phrases percentage related to the negative phrases percentage among social phrases of virtual teams was significantly lower (F 1,46 = 6.912 , p = .012 , η2 = 0.131).

H2c – Virtual team's question versus answer discussion orientation is not different than face-to-face team's questions versus answers discussion orientation.

A Pearson Correlation analysis was conducted in order to measure the correlations between the (social and task) processes of the teams and the teams outputs. The analysis was done for the whole sample (N=50), above the experiment condition. A summary of the results is presented in Table 4.

|               | SocVsTask.P | PosVsNeg.P | QuesVsAns.P |
|---------------|-------------|------------|-------------|
| Effectiveness | -.240       | .350*      | -.159       |
| Efficiency    | .095        | -.480**    | .073        |
| Satisfaction  | -.174       | .398**     | -.170       |

\*\* Correlation is significant at the 0.01 level (2-tailed).
\* Correlation is significant at the 0.05 level (2-tailed).

*Table 4: Pearson Correlations between the Mediators and the Output Variables (Overall N=50)*

The Pearson Correlations indicate that among all teams, there is (H3b) a significant positive linear correlation between Positive Vs Negative Social Level and Effectiveness (r = +.350 , p = .013); a significant negative linear correlation between Positive Vs Negative Social Level and Efficiency (r = -.480 , p < .001); and a significant positive linear correlation between Positive Vs Negative Social Level and Satisfaction (r = +.398 , p = .004). Correlations between Social Vs Task Level and team output (H3a) were not found. Correlations between Questions Vs Task Answers Task Level and team output (H3c) were not found either.

Since the focus of this study is virtual teams, we also performed a Pearson Correlations for that specific experiment condition, namely the virtual teams. A summary of the results is presented in Table 5.

|               | SocVsTask.P |       | PosVsNeg.P |         | QuesVsAns.P |       |
|---------------|-------------|-------|------------|---------|-------------|-------|
|               | virtual     | f-t-f | virtual    | f-t-f   | virtual     | f-t-f |
| Effectiveness | -.349       | -.105 | .478(*)    | .141    | -.115       | -.225 |
| Efficiency    | .324        | .297  | -.303      | -.442(*)| .162        | -.110 |
| Satisfaction  | -.501(**)   | .033  | .453(*)    | -.270   | -.171       | -.228 |

\*\* Correlation is significant at the 0.01 level (2-tailed).
\* Correlation is significant at the 0.05 level (2-tailed).

*Table 5: Pearson Correlations between the Mediators and the Output Variables for virtual teams (N=26) and face-to-face teams (N=24)*

The Pearson Correlations indicate that among virtual teams there is (H3a) a significant negative linear correlation between Social Vs Task Level and Satisfaction (r = -.501 , p = .009); (H3b) a significant positive linear correlation between Social Vs Task Level and Effectiveness (r = .478 , p = .014); and a significant positive linear correlation between Positive Vs Negative Social Level and Satisfaction (r = +.453 , p = .020). Correlations between Questions Vs Answers Task Level and team output (H3c) were not found either.

Among face-to-face teams there is (H3b) a significant negative linear correlation between Positive Vs Negative Social Level and Efficiency (r = -.442 , p = .031). Correlations between Social Vs Task Level and team output (H3a) and Questions Vs Answers Task Level and team output (H3c) were not found.

## INTERIM DISCUSSION

The effectiveness of the virtual teams was not different from the effectiveness of the face-to-face teams, yet: the two other outcomes of the virtual teams were different, virtual team members were less satisfied than face-to-face team members and less efficient. This makes the controversial previous findings reasonable: on the one hand, it isn't necessary anymore to bring team members together to get their best work (Majchrzak et al., 2004) since the distance and the communication type doesn't decrease the team absolute output (effectiveness); on the other hand, the virtual work is much more time consuming and does decrease the team members satisfaction; Implying that the virtual teams are less successful than face-to-face teams on most outcome measures (Denton, 2006; Potter & Balthazard, 2002).

The virtuallity yields a significant effect on the social aspect of the virtual team's discussion: The social aspect is negative than the social aspect of the face-to-face team's discussion, as hypothesized. This evidence is also consistent with previous studies, such as Cramton & Webber (2005) insights about technology tools limitation of information, and the increase in hostility and aggression of the communicators.

Yet, there were no findings of any differences between virtual and face-to-face teams referring to the task versus social discussion orientation, nor referring to the questions versus answers task discussion orientation.

In the light of absent virtuallity effects on task versus social discussion orientation and on questions versus answers discussion orientation, its significant effect on the social part of the virtual team's discussion is even more powerful. The media mainly affects the social context of the environment. Other differences are chain reactions.

The correlation between the (social and task) processes of the teams and the teams output, over the entire sample, demonstrates a significant power of the Positive Vs Negative Social Level of the team discussion. Positive (versus negative) discussion orientation is translated into enhanced effectiveness, member satisfaction, and efficiency (it takes less time to carry out the task); and vice versa: Negative discussion orientation is translated into reduced effectiveness, member satisfaction, and efficiency. As stated before, the way virtual teams manage internal conflict is a crucial factor in their success (Guerra et al., 2005; Montoya-Weiss, Massey & Song, 2001). Therefore, the positive social orientation of the team discussion is important not only as an affected variable, but as an influence variable as well.

A significant correlation between social versus task discussion orientation and the teams' output was not found, nor a significant correlation between question versus answer discussion orientation and the teams' output. This means that social versus task discussion orientation and question versus answer discussion orientation are two variables that are not affected by the virtuallity level and also does not influence the teams' output.

The correlation under the specific experiment condition of virtual teams partially duplicates the findings shown regarding the Positive Vs Negative Social Level of the team discussion over the entire sample. Positive (versus negative) discussion orientation is translated into enhanced effectiveness and members' satisfaction. The efficiency of virtual teams is not affected by the Positive Vs Negative Social Level. Yet, for face-to-face teams, positive (versus negative) discussion orientation is translated only into enhanced efficiency (it takes less time to carry out the task) and have no correlation with the team effectiveness and the members' satisfaction.

Another significant impact was found only for the virtual teams, which is a negative correlation between social versus task discussion orientation the team members' satisfaction. This finding indicates that the more the discussion is socially oriented, the members' satisfaction decreases, and the more the discussion is task oriented, the members' satisfaction increases. The same correlation was not found among the face-to-face teams. It implies that though the social part of the discussion of virtual team should be positive oriented, it (the social part vs. the task part) is expected to be short. The team members are willing to have a task oriented discussion, and not to spend much sources on social talk. If they find they do spend sources on social talk, they lose satisfaction.